\documentclass{emulateapj}
\usepackage{natbib}
\usepackage{epsfig}

\newcommand{\hi}{\mbox{\rm \ion{H}{1}}}

\newcommand{\htwo}{\mbox{\rm H$_2$}}

\newcommand{\acounits}{\mbox{M$_\odot$ pc$^{-2}$ (K km s$^{-1}$)$^{-1}$}}

\newcommand{\sigmass}{\mbox{$\left< \Sigma \right>^{\rm M}$}}
\newcommand{\sigarea}{\mbox{$\left< \Sigma \right>^{\rm A}$}}

\shorttitle{Variable Sub-Kpc Clumping in ISM Maps}
\shortauthors{Leroy et al.}

\begin{document}

\slugcomment{Accepted for Publication in the Astrophysical Journal Letters} 
\title{Clumping and the Interpretation of kpc-Scale Maps of the Interstellar Medium: Smooth \hi\ and Clumpy, Variable H$_2$ Surface Density}

\author{Adam K. Leroy\altaffilmark{1}, 
Cheoljong Lee\altaffilmark{2}, 
Andreas Schruba\altaffilmark{3}, 
Alberto Bolatto\altaffilmark{4},  
Annie Hughes\altaffilmark{5}, 
J\'er\^ome Pety\altaffilmark{6,7},
Karin Sandstrom\altaffilmark{5}, 
Eva Schinnerer\altaffilmark{5}, 
Fabian Walter\altaffilmark{5}}

\altaffiltext{1}{National Radio Astronomy Observtory, 520 Edgemont Road, Charlottesville, VA 22903, USA}
\altaffiltext{2}{Department of Astronomy, University of Virginia, 530 McCormick Road, Charlottesville, VA 22904}
\altaffiltext{3}{California Institute for Technology, 1200 E California Blvd, Pasadena, CA 91125}
\altaffiltext{4}{Department of Astronomy, University of Maryland, College Park, MD, USA}
\altaffiltext{5}{Max Planck Institute f\"ur Astronomie, K\"onigstuhl 17, 69117, Heidelberg, Germany}
\altaffiltext{6}{Institut de Radioastronomie Millim\'etrique, 300 Rue de la Piscine, F-38406 Saint Martin d'H\`eres, France}
\altaffiltext{7}{Observatoire de Paris, 61 Avenue de l'Observatoire, F-75014 Paris, France.}

\begin{abstract}
Many recent models consider the structure of individual interstellar medium (ISM) clouds 
as a way to explain observations of large parts of galaxies. To compare such models to observations, 
one must understand how to translate between surface densities observed averaging over large 
($\sim$ kpc) scales and surface densities on the scale of individual clouds ($\sim$ pc scale), which 
are treated by models. We define a ``clumping factor'' that captures this translation as the ratio of the 
mass-weighted surface density, which is often the quantity of physical interest, to the area-weighted 
surface density, which is observed. We use high spatial resolution (sub-kpc) maps of CO and \hi\ 
emission from nearby galaxies to measure the clumping factor of both atomic and molecular gas. The 
molecular and atomic ISM exhibit dramatically different degrees of clumping. As a result, the ratio 
H$_2$/\hi\ measured at $\sim$ kpc resolution cannot be trivially interpreted as a cloud-scale ratio of 
surface densities. \hi\ emission appears very smooth, with a clumping factor of only $\sim 1.3$. Based 
on the scarce and heterogeneous high resolution data available, CO emission is far more clumped with 
a widely variable clumping factor, median $\sim 7$ for our heterogeneous data. Our measurements do not provide evidence for a 
universal mass-weighted surface density of molecular gas, but also cannot conclusively 
rule out such a scenario.  We suggest that a more sophisticated treatment of molecular ISM structure, one 
informed by high spatial resolution CO maps, is needed to link cloud-scale models to kpc-scale observations of galaxies.
\end{abstract}

\keywords{}

\section{Clumping and Surface Densities in ISM Maps}
\label{sec:intro}

Observations of atomic and molecular gas now achieve spatial resolution of several hundred parsecs 
to a few kiloparsecs in large samples of nearby galaxies \citep[e.g.][]{HELFER03,WALTER08,LEROY09} or even 
small sets of high redshift galaxies \citep{HODGE12,TACCONI12}. Such observations isolate key physical 
conditions such as stellar surface density, metallicity, or the interstellar radiation field. 
However, with a few exceptions, these observations still do not resolve individual clouds of atomic and molecular gas, 
which are often considered to be the fundamental units of the interstellar medium (ISM). 

The interpretation of these observations often utilizes predictions from models that treat the surface density of gas {\em on the scale of individual clouds}. 
For example the models of \citet{KRUMHOLZ09A,KRUMHOLZ09B}, \citet{WOLFIRE10}, \citet{FELDMANN12}, and
 \citet{NARAYANAN12} all consider the structure of individual photodissociation regions or atomic-molecular complexes to
 explain observations on the scale of galaxies. In these models, the surface density of an individual cloud represents a key
parameter, often because it indicates the degree of shielding from the ambient radiation field. Because these models focus on cloud structure
the mapping between the readily observed average, or ``area-weighted,'' surface density at $\sim$ kpc resolution 
and the cloud-scale, ``mass-weighted,'' surface density represents an essential component of comparing observations and theory. This mapping is often 
referred to as ``clumping'' and quantified via a ``clumping factor.''

For the most part the adopted clumping factors represent guesses informed by our coarse knowledge of ISM structure 
and giant molecular clouds (GMCs) but not directly based on observations. However, this factor can also be directly measured 
from high spatial resolution data. In this letter we collect a large set of observations to measure the relationship 
between the surface density of the ISM averaged over large ($\sim$ kpc) scales and the ``true'' small 
scale surface density. We consider both atomic (\hi ) and molecular (\htwo , traced by CO) gas and discuss the implications of 
our calculation for the comparison to models.

We cast this discussion in terms of three quantities: the {\em mass-weighted} average surface density, \sigmass , the {\em area-weighted} average surface density, \sigarea , and a clumping factor, $c$, relating the two. The {\em mass-weighted} average surface density is

\begin{equation}
\label{eq:iwt}
\sigmass = \frac{\int_A \Sigma \times \Sigma~dA }{\int_A \Sigma~dA} = \frac{ \int_A \Sigma^2~dA}{\int_A \Sigma~dA}
\end{equation}

\noindent where $\Sigma$ is the true gas mass surface density along a line of sight and 
the integral occurs over some area element $A$. Then the denominator is simply the sum of 
gas in that area and the calculation returns the mass-weighted average surface density over the 
area. That is \sigmass\ is the column density at which most mass exists. Contrast this quantity with 
what is observed by a telescope for which a resolution element has size $A$,

\begin{equation}\label{eq:awt}
\sigarea = \frac{\int_A \Sigma~dA}{\int_A~dA}~.
\end{equation}

\noindent That is, the telescope observes the area-weighted average surface density within the beam. 

These two quantities, \sigmass\ and \sigarea , will be the same for a smooth medium. They differ 
for a clumpy medium with most of the mass in small, high column density regions spread over large, 
low column density areas. In this case, \sigarea\ may be much lower than \sigmass . We 
define a {\em clumping factor}, $c$, to quantify this distinction as:

\begin{equation}
\label{eq:clumping}
c \equiv \frac{\sigmass}{\sigarea}~.
\end{equation}

\noindent $c$ will be high for a clumpy, inhomogeneous medium and low for a smooth medium. It will 
never fall below unity. In practice, \sigmass\ will be derived at finite resolution, so that $c$ could be more 
rigorously defined as $c_{a}^{b}$, the clumping factor calculated at final resolution $b$ with \sigmass\ derived 
from data with intrinsic resolution $a$. In this paper, $b$ will always be 1~kpc; $a$ will vary from data set to data set.

The clumping factor, $c$, \sigarea , and \sigmass\ give us a formalism to ask several questions related to the structure of the ISM and the interpretation of $\sim$ kpc resolution elements: 

\begin{itemize}
\item How does the mass-weighted $\sigmass$ relate to the area-weighted, observable $\sigarea$ for kpc-resolution measurements of atomic and molecular gas in galaxies? What are typical clumping factors?
\item Is clumping the same for atomic and molecular gas, so that the H$_2$-to-\hi\ ratio at large scales may be readily interpreted in terms of cloud structure?
\item Can one reliably predict the surface density of individual regions --- relevant to PDR and cloud structure calculations --- from coarse resolution measurements? 
\end{itemize}

\begin{figure}
\plotone{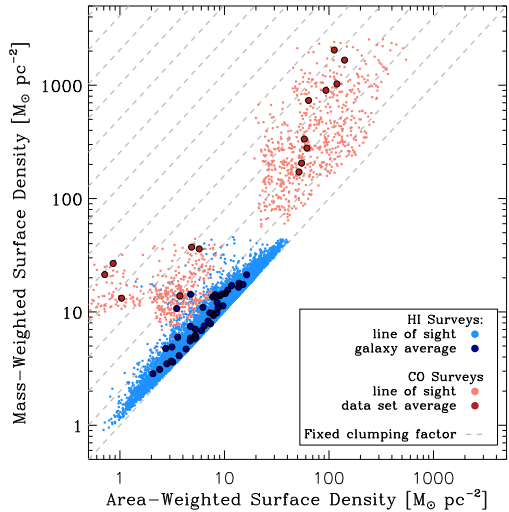}
\caption{Mass-weighted surface density at 1-kpc resolution ($y$-axis), Equation \ref{eq:iwt}, as a function of area-weighted 
surface density, Equation \ref{eq:awt}, at the same resolution ($x$-axis) --- note that no inclination corrections are applied, 
broadening the spread of apparent column densities. That is, the true surface density from which most emission arises as a 
function of the column density that would be measured at 1~kpc resolution. Blue points show \hi\ data, red points show \htwo\ estimates from CO emission. Light points show individual lines of 
sight, dark points show averages for whole data sets. Gray lines show fixed clumping factors spaced by a factor of 2. \hi\ shows 
good tracking between \sigmass\ and \sigarea\ with very low clumping factors, seldom above a factor of two. Conversely, CO exhibits a high degree of 
clumping, almost never less than a factor of two but often more than a factor of ten and a scattered, non-universal relation 
between \sigmass\ and \sigarea .}
\label{fig:col_vs_col} 
\end{figure}

\begin{figure}
\plotone{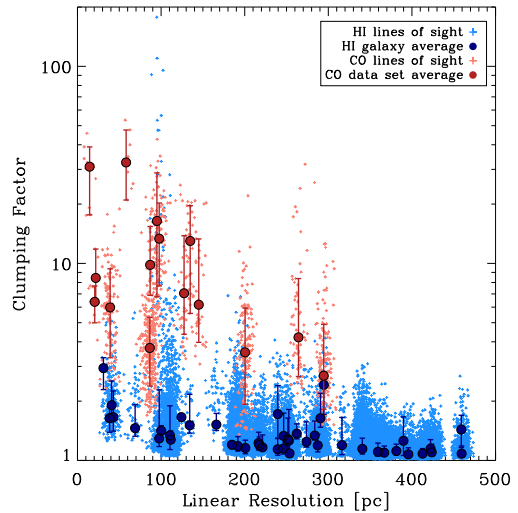}
\caption{Clumping factor ($y$-axis), Equation \ref{eq:clumping}, as a function of the linear resolution of the data used to derive
the mass-weighted surface density, \sigmass , i.e., the native resolution of the data. The color scheme follows Figure \ref{fig:col_vs_col}.
Again, \htwo\ traced by CO appears much more strongly clumped than \hi , exhibiting a wide range of clumping factors and tending
to show a high degree of clumping. \hi , by contrast, appears remarkably smooth, seldom exceeding clumping factors of two even at
high spatial resolution. As in Figure \ref{fig:col_vs_col}, the difference in structure of the atomic and molecular medium and the 
variable clumpiness of the molecular medium are clearly evident.}
\label{fig:clumping} 
\end{figure}

\section{Data and Calculations}
\label{sec:data}

We assemble all readily available high spatial resolution ($\lesssim 500$~pc) CO and \hi\ maps of nearby galaxies and
use these to calculate \sigmass , \sigarea , and $c$. We make use of three recent \hi\ surveys of nearby galaxies: 
THINGS \citep{WALTER08}, LITTLE THINGS \citep{HUNTER12}, and VLA ANGST \citep{OTT12}. We supplement these with a 
collection of \hi\ data obtained to complement the HERACLES CO survey \citep[presented in][]{LEROY12,SCHRUBA11,SANDSTROM12} 
and WSRT maps of M33 \citep{DEUL87} and M31 \citep{BRINKS84}. Whenever possible, we use the naturally weighted data. We 
include all galaxies from these surveys that have linear resolution better than 500~pc and inclination less than $50\arcdeg$ (we 
except M31 and M33 from the inclination requirement). For the \hi\ calculation we consider only regions inside $r_{25}$ 
with column densities $N ({\rm H}) > 10^{20}$~cm$^{-2}$. We use the integrated intensity maps provided by each survey in its data release.

High spatial resolution CO data remains harder to come by than high resolution \hi\ data because nearby dwarf galaxies tend
to be faint in CO emission and the sensitivity of mm-wave telescopes has been limited before ALMA. This
scarcity leads us to assemble a heterogeneous collection of high resolution CO. This includes the 
MAGMA \citep{WONG11} and NANTEN \citep{FUKUI99} surveys of the LMC, the IRAM 30-m
survey of CO in M31 \citep{NIETEN06}, the combined BIMA and FCRAO survey of M33 \citep{ROSOLOWSKY07}, ALMA science verification data
on the Antennae galaxies, and a handful of the brightest and nearest galaxies from BIMA SONG \citep[NGC 2903, 3627, 5194, and 6946]{HELFER03}.
We supplement these with two new datasets: high resolution CARMA mapping of select fields in M31 (PI: A. Schruba, Schruba et al. in prep.) and 
the Plateau de Bure Arcsecond Whirlpool Survey (PAWS, Schinnerer et al. submitted, Pety et al. accepted)
of M51. Except for the ALMA Antennae data, all of these data sets target the CO $J=1\rightarrow0$ line and include 
(sometimes exclusively) short-spacing data; the Antennae data target the CO $J=3\rightarrow2$ and 
CO $J=2\rightarrow1$ transitions\footnote{For purposes of calculating surface densities, we assume these lines to be thermalized. In actuality, 
they are likely somewhat subthermal but uncertainty is likely offset by a somewhat lower $\alpha_{\rm CO}$ in the Antennae. In any case, these 
conversion factors effectively divide out when calculating $c$.}. Note that we have multiple data sets on several galaxies (M31, the LMC, M51) and treat each data set, rather than galaxy, 
as a separate measurement. For comparison, we also calculate $c$ from the composite CO survey of the Milky Way by \citet{DAME01}, considering only intermediate latitude ($30\arcdeg > |b| > 5\arcdeg$) gas. We smooth their data with a $1.25\arcdeg$ kernel to minimize sampling issues and integrate only over areas covered by the surveys.

{\em Calculating Moment Zero Maps:} Because of the $\Sigma^2$ term in Equation \ref{eq:iwt}, \sigmass\ is not robust to
the inclusion of noise in the calculation. We must therefore mask the data before carrying out our calculations. This is mostly an issue for the 
CO data, as the integrated intensity maps provided by the \hi\ surveys have sufficient S/N for our purposes. For the CO, we create new masks and 
re-derive integrated intensity maps for each data set. Typically we estimate the noise from
the empty regions of the cube, identify a core of high significance emission, often two successive channels with $S/N > 5$ and expand this high significance core
to include fainter but still significant emission. We integrate the masked data cube to produce a moment 0 map, which we use in further analysis. For the LMC maps
from MAGMA and NANTEN and the M33 map, the signal to noise at the native resolution is too low to yield a high quality masked integrated intensity map. Therefore we convolve the 
data to a slightly worse resolution before masking and any analysis. 

This exercise produces maps well-suited to derive \sigmass , but the process of masking at the native resolution
 does remove the possibility of picking up any contribution from a low S/N diffuse component (e.g., Pety et al., accepted).
 Fundamentally, this is a limitation of the data themselves and 
 future, more sensitive surveys capable of detecting diffuse emission over individual lines of sight will improve this situation and quantify the contribution of
 faint, pervasive CO emission to the total molecular gas budget\footnote{This effect matters but does not appear to dominate our results. For example, if we add a pervasive 
 CO component to the PAWS M51 data with magnitude $\sim 0.5$ times our sensitivity --- an aggressive scenario --- then $c$ drops from $\approx 6$ to $\approx 5.1$ over the region
 that we consider. Fainter regions, which we avoid, will be more affected.}.
 
 When sampling the CO emission, we restrict ourselves to areas that include significant emission within the mask. Roughly, our criterion is that in the map smoothed to
 1~kpc resolution, the average brightness is such that we could have detected that line of sight at the original, higher resolution. That is, we consider areas where our sensitivity
 at high resolution is sufficient to detect the average brightness. This allows us to avoid ``edge'' or ``clipping'' effects in which only one small patch of bright emission is included in the
 beam, leading to high \sigmass\ but low \sigarea  . Because these ``edges'' are mostly present (within $r_{25}$) in the CO and not \hi\ maps, including them would make our conclusions more extreme.

{\em Deriving \sigmass , \sigarea , and $c$:} We assume that 21-cm and CO emission linearly trace the surface density of atomic and 
molecular gas, i.e., $\Sigma \propto I$, and calculate \sigmass\ and \sigarea\ following Equations \ref{eq:iwt} and \ref{eq:awt}. We convert from intensity to surface density adopting a 
fixed $\alpha_{\rm CO} = 4.35$~\acounits\ and $N \left( \hi \right)~\left[ {\rm cm}^{-2} \right] = 1.823 \times 10^{18} I_{\rm HI}~\left[{\rm K~km~s^{-1}}\right]$. 
Note that these factors divide out when calculating $c$.

To calculate \sigarea , we smooth from the native resolution to 1~kpc using a normalized Gaussian kernel. To calculate \sigmass , 
we calculate $\Sigma^2$ at the native resolution, convolve this map to 1~kpc resolution using a normalized Gaussian kernel, and then 
divide that map by \sigarea\ following Equation \ref{eq:iwt}. We record \sigmass , \sigarea , and the clumping factor, $c$, for a 
hexagonally-spaced set of Nyquist-sampled (at 1~kpc resolution) points.

After this exercise, we have $\approx 50,000$ data points from 46 galaxies for \hi\ and $\approx 1,000$ data points from 15 data 
sets in 8 galaxies for CO. As these numbers make clear, CO data represent the limiting reagent in this calculation, though thanks to 
ALMA their prospect for short-term improvement is excellent .

\section{Results}
\label{sec:results}

\begin{deluxetable}{lccc}
\tablecaption{Clumping Factors for ISM Maps} 

\tablehead{ 
\colhead{Data Set} & 
\colhead{res.} & 
\colhead{$\left< c \right>$} &
\\
\colhead{} & 
\colhead{[pc]} &
\colhead{}
\\
\colhead{(1)} & 
\colhead{(2)} & 
\colhead{(3)}
}

\startdata
CO Data & & \\
\hline 
M31 IRAM 30-m & 87 & $3.7_{-1.3}^{+1.6}$ \\
M31 CARMA ``Brick 9'' & 22 & $8.4_{-2.0}^{+3.4}$ \\
M31 CARMA ``Brick 15'' & 21 & $6.4_{-1.4}^{+1.3}$ \\
M33 BIMA+FCRAO & 98 & $13_{-5.7}^{+6.9}$ \\
LMC NANTEN & 58 & $33_{-12}^{+15}$ \\
LMC MAGMA & 15 & $31_{-13}^{+8.0}$ \\
M51 PAWS & 39 & $6.0_{-2.7}^{+3.4}$ \\
NGC 2903 BIMA SONG & 264 & $4.2_{-1.5}^{+4.2}$ \\
NGC 3627 BIMA SONG & 295 & $2.7_{-0.9}^{+2.2}$ \\
M51 BIMA SONG & 200 & $3.5_{-1.6}^{+2.4}$ \\
NGC 6946 BIMA SONG & 145 & $6.2_{-2.2}^{+7.1}$ \\
Antennae CO(2-1) North & 135 & $13_{-7.4}^{+6.7}$ \\
Antennae CO(2-1) South & 127 & $7.1_{-2.7}^{+6.8}$ \\
Antennae CO(3-2) North & 95 & $7.1_{-9.6}^{+12.4}$ \\
Antennae CO(3-2) South & 87 & $9.8_{-2.9}^{+5.6}$ \\
Local Milky Way ($30\arcdeg > \left|b\right| > 5\arcdeg$) & \nodata & $\approx 6$ \\
\hline
\\
HI ensemble & & $1.26_{-0.15}^{+0.46}$ \\
... 0--250 pc resolution & & $1.34_{-0.17}^{+0.32}$ \\
... 250--500 pc resolution & & $1.19_{-0.11}^{+0.18}$ \\
\enddata
\label{tab:clumping}
\end{deluxetable}

Figures \ref{fig:col_vs_col} and \ref{fig:clumping} and Table \ref{tab:clumping} report our results. Figure \ref{fig:col_vs_col} shows \sigmass\ as a 
function of \sigarea\  for CO (red) and \hi\ (blue) data. Figure \ref{fig:clumping} plots the clumping factor, $c$, as a function of the linear resolution of the 
original data set used to calculate \sigmass . In both figures, light points show individual lines of 
sight and dark, solid points show averages for whole data sets. Error bars  on the whole galaxy points in Figure \ref{fig:clumping} show the $1\sigma$ range for that data set. 
Table \ref{tab:clumping} reports the native resolution (after convolution to increase the S/N in M33 and the LMC) and median clumping 
factor with $1\sigma$ range for each CO data set. We report results for the ensemble of \hi\ data, which Figure \ref{fig:col_vs_col} and \ref{fig:clumping} demonstrate to be uniform.

These figures illustrate three points:

\begin{enumerate}
\item {\em \hi\ and \htwo\ (traced by CO) exhibit different clumping factors.} Both Figure \ref{fig:col_vs_col} and Figure \ref{fig:clumping} show
that essentially {\em all} of our CO data is more highly clumped than {\em all} of our \hi\ data. The median clumping factor for a CO data set is $c=7$, while the 
median clumping factor for an \hi\ data set is $c=1.26$. The specific cases of M33 and M31 illustrate this point cleanly. The M31 IRAM CO map shows 
clumping factor $\approx 4$; the M33 BIMA+FCRAO map shows clumping factor $\approx 13$. Both the M31 and M33 \hi\ maps show clumping factor $\approx 1.3$. 

As a direct result, a ratio of H$_2$-to-\hi\ surface densities obtained at large scales does not translate
trivially into a ratio of surface densities at small scales. The assumption that the large scale surface density in galaxies reflects the small scale surface density
in the same way for \htwo\ and \hi\ underlies the application of the \citet{KRUMHOLZ09A} model to explain \htwo -to-\hi\ ratios in galaxies. Though the
physics of the model appear to apply successfully to individual clouds or regions \citep{BOLATTO11,LEE12}, we suggest that
more than a single ``clumping'' factor is necessary to make a rigorous comparison of the model to observations of large parts of galaxies.

Our calculation does not invalidate the kpc-resolution ratio of $\Sigma_{\rm H2}/\Sigma_{\rm HI}$ as an interesting measurement. It simply suggests
that this be viewed as a measure of mass balance among ISM phases over a large area and not indicative of small-scale ISM structure.

\item {\em \hi\ is very smooth.} This conclusion leaps out of both figures. Even with high linear resolution, \hi\ column density maps remain smooth 
and only weakly clumped. This can be explained by most 21-cm emission originating not from 
clumped bound clouds, but a diffuse medium with a high volume filling factor. The \hi\ clumping factor does depend weakly on scale, a reasonable functional form is 
$c = 392~\left(l_{\rm pc} + 100\right)^{-1.27} + 1$, where $l_{\rm pc}$ is the (FWHM) linear resolution in parsecs.

\item {\em CO is clumpy with a wide range of $c$, making it hard to predict \sigmass\ from \sigarea .} In contrast to \hi , \htwo\ traced by CO emission
appears clumpy with a wide range of $c$. The median among all data sets is $\approx 7$ with a factor of $2$--$3$ rms scatter among
measurements. This is also close to the value that we estimate for the Solar Neighborhood from intermediate latitude gas, but we caution that we expect $c$ to change with
improving native resolution of the data. Because we consider only bright regions, this represents a conservative estimate; the edges and faint regions that we exclude
tend to have high $c$. The molecule-poor systems (M33 and the LMC) in the sample show the highest $c$, perhaps because they contain more isolated clouds and perhaps
because their somewhat low metallicities lead any diffuse H$_2$ component to emit less in CO (below some metallicity the clumping of CO emission and $\Sigma_{\rm H2}$ will 
dramatically diverge). \\
\end{enumerate}

\noindent We also note two less secure points implied by the data but requiring more aggressive assumptions about the CO-to-H$_2$ conversion factor:

\begin{enumerate}
\setcounter{enumi}{3}

\item {\em The total (\htwo $+$ \hi ) clumping factor must vary significantly among and within galaxies.} We can only calculate the 
clumping factor for the total (\htwo $+$ \hi ) gas in M31, M33, and the LMC. In each case the median $c$ is low 
($\sim 1.3$, $\sim 1.3$, and $\sim 1.5$), resembling that of the \hi . This is not surprising because for fixed $\alpha_{\rm CO}$,
the \hi\ mass exceeds the H$_2$ mass in these galaxies by more than an order of magnitude, with H$_2$ making up most
of the gas along only a small fraction of the lines of sight at our resolution (M31 is more molecule-rich than the other two, but still \hi\ dominated). Generally,
we expect that across most of the area in dwarf galaxies, which tend to be low-metallicity and \hi -dominated, $c$ will resemble the $\sim 1.3$ that 
we measure for \hi . The outer parts of most spirals also tend to be \hi\ dominated and should show similar values, 
while in the molecule-rich central parts of actively star-forming galaxies the values will more closely resemble the higher $c$ that we find for M51, NGC 2903, NGC 3627, and
NGC 6946.

\item {\em CO exhibits a wide range of \sigmass .} In contrast to the common assumption that CO emerges from a population of fixed surface 
density clouds, the CO data in Figure \ref{fig:col_vs_col} span two orders of magnitude in \sigmass . The figure provides no 
good evidence that CO emerges from a fixed \sigmass\ at high spatial resolution. In fact, the highest resolution data sets
span roughly an order of magnitude (for fixed $\alpha_{\rm CO}$) in \sigmass\ from the LMC ($\sim 25$~M$_\odot$~pc$^{-2}$) to M51 ($\sim 330$~M$_\odot$~pc$^{-2}$). 
The spread in \sigmass\ may arise from sources other than the cloud scale surface density: variations in inclination and the conversion factor, 
superposition of fixed surface density clouds, the convolution of bound clouds with a diffuse background, or a lack of spatial resolution matched to individual clouds. However, 
our best guess is that the large range in apparent \sigmass\ visible in Figure \ref{fig:col_vs_col} in fact reflects a systematic dependence of cloud surface density on environment. 
This reinforces the thorough analysis of Hughes et al. (accepted), who compare CO 
maps of M51 (PAWS), the LMC (MAGMA), and M33 and conclusively demonstrate fundamental differences among the volume and surface density PDFs among and 
within the three galaxies \citep[see also the spread in Milky Way $\Sigma$ discussed by][]{BOLATTO13}. 

\end{enumerate}

\noindent Our calculations shows that the structure of the molecular and atomic ISM are more
complex than has been assumed while vetting recent models. We suggest the ``clumping factor'' 
approach defined in \S \ref{sec:intro} to quantify this structure and aid interpretation of lower resolution observations. 
With ALMA now able to easily obtain high resolution, high sensitivity ISM maps we expect such calculations to
be feasible in many systems over the coming years.

\acknowledgments We thank the referee for a constructive report and Scott Schnee and Mark Krumholz for feedback on drafts. 
We acknowledge the BIMA SONG, LITTLE THINGS, and VLA ANGST collaborations for making their data public. 
We thank IRAM for making the moment 0 map of M31 public. This paper makes use of the following ALMA data: ADS/JAO.ALMA\#2011.0.00003.SV. ALMA is a partnership of ESO (representing its member states), NSF (USA) and NINS (Japan), together with NRC (Canada) and NSC and ASIAA (Taiwan), in cooperation with the Republic of Chile. The Joint ALMA Observatory is operated by ESO, AUI/NRAO and NAOJ. The National Radio Astronomy Observatory is a facility of the National
Science Foundation operated under cooperative agreement by Associated
Universities, Inc. AB acknowledges partial support from grants 
NSF AST-0838178, NSF AST-0955836, and a Cottrell Scholar award from the Research Corporation for Science Advancement. 
KS is supported by a Marie Curie International Incoming fellowship. 
AH acknowledges funding from the Deutsche Forschungsgemeinschaft via grants SCHI 536/5-1 and SCHI 536/7-1 as part of the priority program
SPP 1573 'ISM-SPP: Physics of the Interstellar Medium.' JP was partially funded by the grant ANR-09-BLAN-0231-01 from the French
{\it Agence Nationale de la Recherche} as part of the SCHISM project.

\end{document}